# Public Sector Sustainable Energy Scheduler – A Blockchain and IoT Integrated System

Renan Lima Baima[1*], Iván Abellán Álvarez[1], Ivan Pavić[1], Emanuela Podda[1]

[1]Interdisciplinary Centre for Security, Reliability and Trust - SnT, University of Luxembourg

{renan.limabaima; ivan.abellan; ivan.pavic; emanuela.podda}@uni.lu

(*Corresponding Author: renan.limabaima@uni.lu)

**ABSTRACT**

In response to the European Commission's aim of cutting carbon emissions by 2050, there is a growing need for cutting-edge solutions to promote low-carbon energy consumption in public infrastructures. This paper introduces a Proof of Concept (PoC) that integrates the transparency and immutability of blockchain and the Internet of Things (IoT) to enhance energy efficiency in tangible government-held public assets, focusing on curbing carbon emissions. Our system design utilizes a forecasting and optimization framework, inscribing the scheduled operations of heat pumps on a public sector blockchain. Registering usage metrics on the blockchain facilitates the verification of energy conservation, allows transparency in public energy consumption, and augments public awareness of energy usage patterns. The system fine-tunes the operations of electric heat pumps, prioritizing their use during low-carbon emission periods in power systems occurring during high renewable energy generations. Adaptive temperature configuration and schedules enable energy management in public venues, but blockchains' processing power and latency may represent bottlenecks setting scalability limits. However, the proof-of-concept weakness and other barriers are surpassed by the public sector blockchain advantages, leading to future research and tech innovations to fully exploit the synergies of blockchain and IoT in harnessing sustainable, low-carbon energy in the public domain.

**NONMENCLATURE**

*Abbreviations*

| | |
|---|---|
| APEN | Applied Energy |
| IoT | Internet of Things |
| PoC | Proof-of-Concept |
| PSB | Public Sector Blockchain |
| EHP | Electric Heat Pump |
| IFTTT | If This Then That |

*Symbols*

| | |
|---|---|
| $BLC$ | Building Load Coefficient |
| $\min_{ehp\_num}$ | Minimum EHP operational hours |
| $\max_{ehp\_num}$ | Maximum EHP operational hours |
| $RE_{gen}$ | Renewable energy share generation |
| $agg_{gen}$ | Total energy generation |
| $EHP_{hours}$ | Electric heat pump operational hours |

## 1. INTRODUCTION

The critical need to reduce carbon emissions and promote renewable energy generation underscores the global urgency to transition to a sustainable economy [1]. Strategies that favor renewable energy adoption encompass grid infrastructures [2], demand response programs [3], energy storage integration [4], and smart EV charging [5]. Renewables' intermittency and legislative requirements, exemplified by the European Commission's 2050 targets [6–8], heighten the need for inventive, policy-compliant solutions [9].

Emerging technologies, specifically blockchain and the Internet of Things (IoT) offer promising avenues to

---

# This is a paper for 15th International Conference on Applied Energy (ICAE2023), Dec. 3-7, 2023, Doha, Qatar.

enhance sustainable energy management [10,11], especially within the public sector characterized by high-energy demand facilities [12]. Such combined technologies can facilitate more transparent and democratic decision-making in the public sector [13,14], in line with the European Strategy for Data [15] that introduced a cross-sectoral legal framework supporting technologies focusing on open data-driven solutions for the benefit of citizens and businesses, among which are solutions for increasing sustainability and energy efficiency. To this end, we propose an architecture integrating blockchain, IoT devices' control, and predictive low-carbon energy availability, explicitly focusing on Electric Heat Pumps (EHP) to maximize renewable share energy consumption. Our system integrates weather predictions from an external data source [16] to identify low-carbon periods, synchronizing energy-intensive activities, like adjusting building temperatures, with these periods to promote sustainable energy consumption in tangible government-owned public assets.

As we developed an e-government application, our research context frames primarily on engineering and policy perspectives, drawing on grey literature to provide a comprehensive view of policy developments on sustainable energy efficiency. This approach positions our work on whether current regulations already cover the issues at hand or if new, creative solutions are needed for problems yet to be regulated. We recognize the importance of research in promoting the adoption of new technologies that improve building energy performance and energy efficiency [17]. The novelty of our work focuses on incentivizing education and information awareness to encourage the adoption of low-carbon energy-saving habits via informative, operational graphs [18,19]. We employ IoT to gauge compliance in public entities, logging data directly into the Public Sector Blockchain (PSB) via smart contracts.

Though our solution is currently functional, specific individual components, like our scheduling algorithm, serve as placeholders, and by making our architectural Proof of Concept (PoC) publicly available, we aim to foster refinement and continuous advancements. Our main goal is to demonstrate the architecture's feasibility, flexibility, and benefits, with individual components being adaptable and modular based on feedback and progress. Given the existing gaps in the literature, our study seeks to answer the question:

*RQ: Can an integrated blockchain and optimized predictive system schedule and operate EHP in tangible government-held public assets to prioritize low-carbon energy consumption?*

By broadening the scope of blockchain implementation beyond e-government solutions [20,21], the system could increase accountability for public spending [22], ultimately benefiting citizens with public resources management. While previous studies have examined the PSB's role in energy usage transparency, sharing information [23], and carbon credit offsetting [24], our approach aims to optimize low-carbon energy consumption for IoT-enabled public assets. This goes beyond simple consumption tracking, considering asset input, scheduling, monitoring, and engagement factors.

The remainder of the paper delves deeper into our approach, methodology, and findings. Section 2 delivers an overview of IoT energy management, automated consumption, PSB, and sustainable optimization. Section 3 outlines the approach and resources for the solution toward the PoC design. Section 4 discusses the heuristic for EHP scheduling. Section 5 showcases PoC outcomes and strategies for optimizing energy in the public sector. Section 6 evaluates PSB operations, IoT data retrieval methods, and broader implications. Lastly, section 7 summarizes the findings and contributions, emphasizing the prospect of employing PSB and IoT for engaging sustainable consumption.

## 2. LITERATURE REVIEW

This section aims to locate our study within the broader literature discussing regulations [22], the role of blockchain in the energy management of IoT devices [2,25], the potential of PSB [26–28], and optimization in smart grids and smart homes [29,30]. We present concepts, identify system gaps, and then focus on the knowledge our research seeks to contribute.

### 2.1 Regulatory framework and sustainable innovation

Complying with the legislative actions promoted at the international and European levels [31], the energy sector is inherently adapting and incorporating innovative technologies, integrating digitalization with the evolution of networks. The shift toward an intelligent paradigm [32] provokes a rethink of the structure in the sector [33], where energy efficiency is crucial. Despite the specific requirements imposed on Member States by Directive 844/2018 [34] – after a legal recast of the 2010 Directive [35] – in 2020, the European Court of Auditors [36] reported a lack of energy efficiency in buildings, recommending more proactiveness in this regard. Consequently, the new European Commission has pledged to apply the "energy efficiency first" principle [37] in developing the new energy strategy, becoming one of the top objectives [38] of the Cohesion policy. Furthermore, the Court of Auditors stressed a need for more transparency in how the EU and Member States report on climate and energy performance [39]. Moreover, they highlight that the EU's contribution to energy efficiency needs to be improved, whatever the sector they operate in [40].

In line with these premises, this paper aims to tailor an innovative solution to address the mentioned criticalities and promote low-carbon energy consumption in tangible government-held public assets. We rely on data availability and benefits from the specificities of two main technologies: PSB and the IoT.

*2.2 Public Sector Blockchain and IoT*

The PSB initiative fosters cross-border collaborations across public and private sectors, delivering a decentralized, transparent platform [27]. It employs a permissioned blockchain system, which means only consortium members can access the data and choose which information to publicize. Via the replicated database, it grants trust, and as it involves only known participants, it allows for more efficient coordinated consensus, eliminating the need for a trusted administrator [41]. Given its predefined participants, the PSB provides a faster latency (i.e., information transfer) while maintaining the same level of data reliability, verifiability, and tamper-proof evidence [27]. Likewise, PSB supports transparent, tamper-proof parameter-based programs executing predefined input functions called smart contracts [41].

European government entities actively examine the blockchain's potential in public sectors to enhance transparency and build trust [26,28]. The primary objective revolves around advancing digital transformation, information storage, and enhancing transparency in public sector decision-making within a context of inherent trust [42].

Despite the recognized benefits of PSB and smart contracts across various applications, such as e-voting, land registration, electronic records [42], copyrights, and potential adoption of PSB by public energy utilities [28], the potential for controlling IoT devices to engage their institutions to participate in low-carbon energy-saving behaviors is mainly unexplored. PSB can be a powerful tool for governments [44] to engage their institutions in sustainable energy consumption while managing IoT devices transparently [43] and ensuring inherent trust. However, our research aims to address particular challenges [43], such as resource limitations inherent to low-powered devices and smart contracts capabilities, especially in managing EHP via a blockchain scheduler.

*2.3 Blockchain IoT and Sustainable Energy Management*

Integrating IoT in energy management represents a step towards a decentralized energy management system. Fundamental studies [45, 46] laid the groundwork for energy exchange and coordination frameworks (e.g., energy trading), proposing the potential of blockchain technology to decentralize and capitalize on energy management between prosumers and consumers [43]. Other research offers decision-support methods for evaluating photovoltaic systems for energy generation, thereby promoting low-carbon consumption [1].

International organizations widely acknowledge the importance of renewable energy prioritization in sustainable energy management systems [47,48]. Existing studies affirm the efficacy of smart home energy management, suggesting the utility of automation and peer-to-peer principles for optimal consumption [29,30]. However, the role of blockchain, specifically PSB, in managing public assets to facilitate sustainable energy consumption still needs to be discovered.

In energy management, blockchain has influenced the use of IoT devices in smart grids [2], decentralized energy trading systems [49,50], motivation for low-carbon consumption in public infrastructure [25], and many other systems [51]. Besides price-based demand response models [52], other incentives, such as the awareness of energy-saving behaviors [18,19] and blockchain-supported carbon trading platforms [24], all contribute to strategically promoting sustainability goals. These technologies provide a roadmap for effective energy management and encourage sustainable practices. Nevertheless, investigations are still needed to explore the capacity of automated scheduling of IoT devices within PSBs for sustainable energy consumption.

## 3. MATERIAL AND METHODS

The inherent trust characteristics of the PSB [27] and the optimized low-carbon energy scheduling [53] support our PoC, serving as a smart contract use case [42]. As further described in section 5, we employed IoT device control [2, 25], optimized for low-carbon scheduling [53], engaging entities via awareness incentives [18,19].

*3.1 Data Collection, Analysis, and Scheduling*

We gathered and processed forecasts for renewable energy production [16] and combined them with information on energy usage from public assets to create reliable insights. The data are fed into the scheduling system that optimizes the energy usage considering public institutions' settings input, such as desired temperature, building construction year, living space, basement availability, and roof insulation for public institutions. Hourly measurements are taken to determine the energy consumption profile, allowing for a comprehensive analysis of the thermal behavior of the heating system and buildings. The profile is established by considering three types of demand: space heating, lighting, and appliances. We customized the required energy for space heat, given occupants' preferences.

*3.2 Optimizing Renewable Energy Consumption*

We utilized an approximation from the literature to optimize renewable energy consumption to guide our PoC [53]. The literature suggests using a generation matching control algorithm for centralized building decisions to operate clusters of EHPs during a mismatch in generation and demand until the difference falls within an acceptable range. Their case study revealed the potential of domestic dwellings' thermal inertia to align with renewable generation, achieving up to 95% self-sufficiency in wind scenarios. Their model, simulating domestic dwellings' load profiles, emphasized demand flexibility, especially in space heating. However, our focus is on general low-carbon availability and the individual control of a single EHP, considering the building structure and its capacity to retain heat [12]. With hourly operational resolution for renewable generation, we aim to increase the utilization of renewable energy generation within the grid [54].

The PoC optimization algorithm determines the optimal operating hours for IoT devices, considering the information from the algorithm preceding the execution of the current one to also consider the current temperature, the building load coefficient (BLC), and the availability of renewable energy in the grid as conditions for the scheduling. We implemented a derived and simplified version of the referenced algorithm [53], supported by the mathematical description discussed in section 4 and data analysis techniques to determine the optimal EHP operating scheduling hours.

*3.3 IoT and Blockchain Integration*

As illustrated in Fig. 1, the PoC integrates smart contracts to register IoT devices' usage into the PSB and pushes into the IoT devices the optimized schedule control. Ethereum's Solidity statically typed programming language was used to develop these smart contracts, while JavaScript handled the backend operations, interfacing with the blockchain via the ethers.js library.

We use the If This Then That (IFTTT)[1] service to control and supervise public EHPs' operations based on renewable energy projections and user input. The IFTTT's applet functionality was leveraged to program webhooks to trigger the IoT devices' operational time via HTTP requests. Despite possible drawbacks, including the system's internet reliance and potential privacy concerns [44], the interoperability and automation capabilities aligned well with our project's engagement goals, making it the most suitable option.

**4. THEORY**

The intersection of comfort and sustainable energy efficiency is most pertinent in indoor temperature regulation. For most households, heating and cooling systems consume up to 55% of the total energy [12]. Our PoC, therefore, employs a heuristic formula to determine EHP operating hours due to its considerable energy footprint. While the architecture can be adapted for other IoT devices within the IFTTT environment, the energy implications of EHPs demand a more tailored approach. Drawing from the utility analysis model [12,53], the formula incorporates the current building temperature, the target temperature, and the $BLC$. The simplified version of this model offers a quick and reasonably accurate estimate, making it apt for a PoC that can undergo iterative refinements.

- Temperature Difference $|T - 20°C|$: Calculates the heating or cooling requirement by contrasting the current temperature with a target (averaged to 20₀C). Based on the concept of building recovery time, which is the time a building takes to recover from its setback to its occupied set point.
- Building Load Coefficient ($BLC$): This variable evaluates energy efficiency by comparing power use to ambient temperature, indicating how effectively the building retains energy.
- Maximum and minimum EHP Operating Hours ($max_{ehp\_num}$ and $min_{ehp\_num}$): The outcome is adjusted according to the minimum and maximum number of hours the EHP must function in a day. Additionally, it is combined with the specific hours the grid has the most significant share of renewables, which optimizes the start times and operation of heating systems during those periods. The heuristic approximation aligns with optimizing heating system operations based on the current and target temperatures and the $BLC$ cap.
- The formula to estimate EHP operation is given by:

$$EHP_{hours} = \max\left(min_{ehp\_num}, \left\lceil max_{ehp\_num} \frac{\alpha \frac{|T - 20°C|}{20°C} + \beta(1 - BLC)}{\alpha + \beta} \right\rceil\right)$$

$$\alpha, \beta, BLC \in [0,1] \text{ and } T \in [0°C, 40°C]$$

$$min_{ehp\_num} \leq max_{ehp\_num} \in \{0,1,\ldots,24\}$$

---
[1] https://ifttt.com/

Where $T$ is the current temperature, BLC is the building load coefficient, $\alpha$, and $\beta$ are the normalization factors, and $max_{ehp\_num}$ and $min_{ehp\_num}$ are the maximum and minimum daily EHP operation hours.

Our system also calculates the share of renewable energy (RE) in total energy generation to promote energy efficiency and sustainability. We generalize in our heuristic, as in total energy generation:

$$RE_{share} = \frac{RE_{gen}}{agg_{gen}}$$

Where $RE_{gen}$ is the renewable energy generation, and $agg_{gen}$ is the total energy generation.

The optimized hours for the EHP to operate to maximize the use of renewable energy is determined by:

$$EHP_{on} = \begin{cases} True, & \text{if } i \in \text{sorted indices of } RE_{share}[-ehp_{hour_{num}}:] \\ False, & \text{otherwise} \end{cases}$$

Finally, the increase in the share of renewable energy when the EHP is operational is calculated by:

$$RE_{share\_increase} = \frac{\frac{1}{EHP_{hours}} \sum_{i=1}^{24} RE_{share}(i) \times EHP_{on}(i)}{\frac{1}{24} \sum_{i=1}^{24} RE_{share}(i)} - 1$$

Where $RE_{share}$ is the share of renewable energy, and $EHP_{on}$ is a binary variable indicating the EHP state.

## 5. RESULT: ARCHITECTURE

Fig. 1 illustrates the architecture of the PoC system, designed to control IoT devices supported by several components, each playing a pivotal role in the system's functionality, processing deeper operations through:

1. **User Interface:** Public sector institutions can set up their environment-centric parameters via the Flask interface, composed of temperature, construction year, dimensions, and basement and roof insulation.
2. **Data Collection and Preprocessing:** Parameters are collected from public entities' inputs and processed through a dataset.
3. **Energy Consumption Prediction:** Based on public entities' inputs and the dataset, the PoC anticipates energy consumption patterns normalized between the 0 and 1 range.
4. **Optimization and Scheduling:** The system interfaces with the external service, acquiring an optimized schedule for the IoT devices per the predicted energy consumption trends. Here, the focus is determining the ideal operation windows to boost energy efficiency and renewable sources.
5. **Real-time Management of IoT Devices:** With the optimized schedule, directives are sent to the IoT devices via the IFTTT platform, ensuring operations during intervals of most energy efficiency.
6. **Performance Metrics:** The system retrieves and presents on the interface three core metrics for increased public entities engagement:
    - Operational status visualized across 24 hours
    - The percentage of renewable energy consumption over the previous day
    - The augmentation of renewable utilization as a system optimization outcome

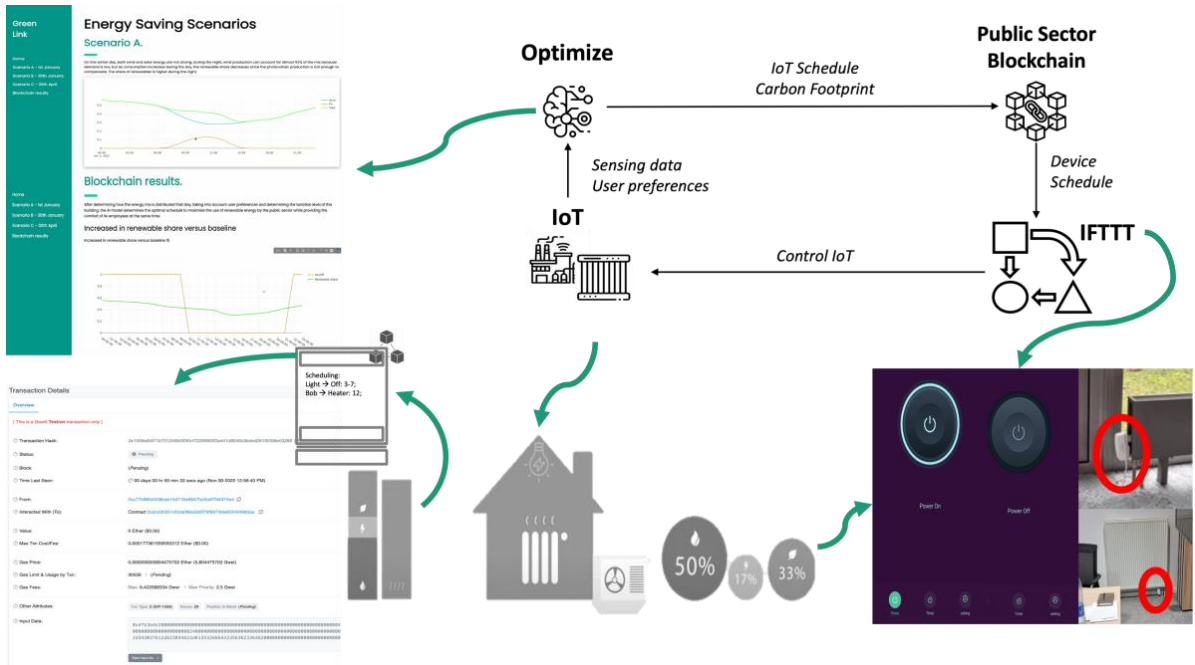

Fig. 1. Overview of the PoC system architecture

## 6. DISCUSSION AND ANALYSIS

During the Data Collection and Preprocessing stage, the PoC underwent scenario testing based on real-world weather data and the precision in controlling IoT devices to validate its effectiveness. As depicted in Fig. 2 on the "Blockchain Results" display, the system is assessed based on three scenarios. We compared each scenario's anticipated EHP operation with renewable energy production on random days in 2022. It displayed a significant increase in renewable shares compared to no scheduler usage, particularly under favorable conditions. The interface then emphasizes this increase in renewable energy proportion relative to the baseline.

These findings illustrate the successful integration of blockchain for transparency and IoT to enhance renewable energy consumption in tangible government-held public assets. This approach, supported by the informative graphs, may promote public entities' engagement and incentivize low-carbon energy consumption behaviors [19]. The PoC signifies gains in energy conservation, presenting a potential solution for reducing $CO_2$ emissions of public sector establishments.

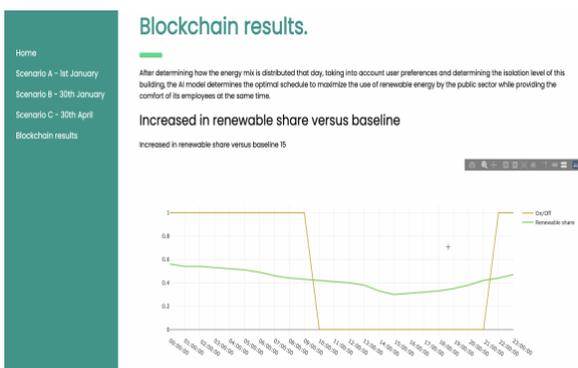

Fig. 2. Visualization of Renewable Energy Consumption in PSB

Still, it is essential to underscore that while the approach encapsulates foundational concepts, its heuristic nature might demand more rigorous validation via diverse datasets or advanced building energy simulation tools.

However, the PoC already reveals inherent limitations. Scalability challenges emerge when considering large-scale, real-world applications, such as processing power constraints for extensive data and IoT devices' energy consumption [43]. The derived heuristic model only captures some of the intricacies of advanced methods known for higher precision [12], like optimal start times and recovery periods [53]. While the metrics embedded in the code offer a comprehensive performance assessment, the limited quantitative data suggests the need for a more detailed evaluation. Nonetheless, the PoC validates the architecture by showcasing potential in augmenting low-carbon energy efficiency for tangible public assets with the help of optimizing weather forecasts and real-time data assimilation, emphasizing the significance of renewable sources. This paves the way for future research to delve deeper into methodologies and complex variables for a refined system proposal.

Upcoming research could prioritize refining the control algorithm, concentrating on power dynamics between renewable generators and the grid. Prior works [55,56] have indicated that this "self-supply" energy balance can be expressed in terms of total energy generation or relative to the overall load. Furthermore, critical metrics, such as the Percentage Dwelling Discomfort (PDD) [53], also factor gauging potential temperature deviations from standard business-as-usual confines of $16.5°C – 21.5°C$. Future efforts could focus on more agile algorithms, predicting day-to-day energy demands, richer data stream integration, and leveraging advanced IoT technology. Despite the challenges, the potential and feasibility of the approach are proven to be a venue for continued research and innovation.

## 7. CONCLUSION

The PoC system demonstrates the architecture's potential to foster optimized IoT device management, aiming to amplify renewable energy consumption engagement within tangible government-owned assets via the PSB framework. By introducing two crucial metrics – renewable share and enhanced renewables usage – our study quantifies and validates the system's impact, offering public entities a clear perspective on their sustainable energy choices. Our systematic comparison of the operation across various scenarios with renewable energy production data from representative days in 2022 revealed a marked 15% increase in renewable energy shares when using the scheduler. This is particularly evident under conditions optimized for renewable energy harnessing compared to our defined baseline of non-optimized operations. This data measures the system's overall performance and offers public entities a transparent understanding of their energy choices' broader sustainable implications.

A standout feature is the synergy between blockchain and IoT technologies, which the PoC showcases as a viable strategy to mitigate carbon emissions in government-held assets. Nevertheless, several obstacles still exist, such as the decentralized data management in blockchain and the proposed incentive structures that might be impractical in resource-constrained countries [57]. While our approach primarily targets specific public infrastructure facets, its broader applicability demands rigorous evaluation of its adaptability across varying sectors, considering their distinct socio-economic and regulatory landscapes and available incentive structures. Recognizing these nuances, future research should also delve into developing regulatory-compliant solutions that balance public interests and foundational rights.

In addressing the central **RQ**, we answered with the PoC's automation awareness incentives preliminary probe into the PSB [18,19], gauging the IoT control to promote transparency and low-carbon behavior in the public sector. This study paves the way for in-depth exploration into incentive mechanisms, stakeholder engagement, and more extensive scale structure scheduling control. It accentuates the importance of comprehensive energy market stakeholder inclusion. Embracing such inclusivity will provide a more rounded view of the strengths and challenges, propelling advancements in sustainable energy practices backed by IoT and incentivized PSB.

## ACKNOWLEDGEMENT

This research was funded by the Luxembourg National Research Fund (FNR) in the FiReSpARX Project, ref. 14783405, the PABLO project ref. 16326754, and by PayPal, PEARL grant reference 13342933/ Gilbert Fridgen. Last, we thank Joaquín Delgado Fernández for his helpful comments on the work and Timothée Hornek for his guidance on the methodology.

## DECLARATION OF INTEREST STATEMENT

The authors declare that they have no known competing financial interests or personal relationships that could have appeared to influence the work reported in this paper. All authors read and approved the final manuscript.